\documentclass[12pt,preprint]{aastex}
\usepackage{color}

\begin{document}

\title{On the e$^+$e$^-$ excesses and the knee of the cosmic ray spectra
--- hints of cosmic rays acceleration at young supernova remnants}

\author{Hong-Bo Hu\altaffilmark{1}, Qiang Yuan\altaffilmark{1},
Bo Wang\altaffilmark{1}, Chao Fan\altaffilmark{2,1},
Jian-Li Zhang\altaffilmark{1} and Xiao-Jun Bi\altaffilmark{1}}

\affil{$^1$Key Laboratory of Particle Astrophysics, Institute of High
Energy Physics, Chinese Academy of Sciences, Beijing 100049, P.R.China\\
$^2$Department of Physics, Shandong University, Jinan 250100, P. R. China}

\begin{abstract}

Supernova remnants have long been regarded as sources of the Galactic
cosmic rays up to petaelectronvolts, but convincing evidence is still
lacking. In this work we explore the common origin of the subtle 
features of the cosmic ray spectra, such as the knee of cosmic ray
spectra and the excesses of electron/positron fluxes recently observed 
by ATIC, H.E.S.S., Fermi-LAT and PAMELA. Numerical calculation shows 
that those features of cosmic ray spectra can be well reproduced in a 
scenario with e$^+$e$^-$ pair production by interactions between high 
energy cosmic rays and background photons in an environment similar to 
the young supernova remnant. The success of such a coherent explanation 
serves in turn as an evidence that at least a portion of cosmic rays 
might be accelerated at young supernova remnants.

\end{abstract}


\maketitle

Ever since the discovery made by \cite{KK1958}, the knee of the cosmic 
ray (CR) spectra observed $\sim4$ PeV has remained a puzzle for half a 
century. The origin of the knee has become a key problem of CR physics, 
as it is closely related to the acceleration of the Galactic CRs. 
There have been extensive studies on the issue (e.g., see the review of 
\cite{2006JPhCS..47...41H} and references therein), however, no consensus 
has been reached yet due to the lack of high precision experimental 
information.

Recently, the ATIC experiment measured the cosmic electron (electron and
positron) spectrum up to several TeV \citep{2008Natur.456..362C} with 
high precision and showed evidence for an obvious excess of flux above
tens of GeV. In addition, the ground based atmospheric Cerenkov
telescope H.E.S.S. and the space satellite Fermi, which are
dedicated to high energy $\gamma$-ray detection, also reported
their results on the electron spectrum measurements
\citep{Fermi:2009zk,2008PhRvL.101z1104A,Aharonian:2009ah}. Despite
some discrepancies between the results of those experiments, 
all of them seem to show the excess of electrons
compared with the background estimation. At the same time, the
PAMELA experiment also published their results on positron fraction
for energies up to $\sim 100$ GeV \citep{2009Natur.458..607A}.
The PAMELA results show clear positron excess, in agreement with the
anomaly observed before by balloon-borne experiments
\citep{1997ApJ...482L.191B,2007PhLB..646..145A}.

It is suggested that the rise of the positron fraction may require
a primary source of e$^+$e$^-$ pairs \citep{2009PhRvD..79b1302S}.
As early as the 1990’s, Aharonian and Atoyan proposed a model involving
the e$^+$e$^-$ pair production mechanism from the interactions between
high energy $\gamma$-rays and optical or ultraviolet radiation in the
vicinity of discrete sources, to explain the electron/positron excesses
from early time balloon-borne experiments \citep{1991JPhG...17.1769A}. 
We think that it might be more probable to explain the lepton excess
by the e$^+$e$^-$ pair production via interactions between CRs and
background photons. Such a process would be much more efficient than 
the $\gamma\gamma$ interaction because the cross sections are comparable, 
while in the former case multiple collisions are allowed.

In addition to the production of e$^+$e$^-$ pairs, the interaction
between nuclei and background radiation will unavoidably change
the spectrum of CRs on the source and may lead to the formation of
the knee. Considering the radiation at the source is at the
optical energy level ($\sim$eV), we expect the pair production to 
occur at $\sim 1$ PeV for protons and a few PeV for Helium, which 
just corresponds to the ``knee'' of CR spectra. In the rest frame 
of CR nuclei, the secondary electrons/positrons have energies around 
MeV, which turn to be about TeV in the laboratory frame. It is striking
to note this is just the energy range where the excesses have been 
observed by ATIC, Fermi and H.E.S.S.. As for the energy budget, the 
energy density of the excess electrons observed by ATIC is about 
$3\times 10^{-5}$ eV cm$^{-3}$ between $0.1$ and $1$ TeV, which is of 
the order of CR energy loss assuming a spectral break from 
$-2.7$ to $-3.1$ at energy $\sim 1$PeV\footnote{Note that this 
estimate is based on the observational spectra. The energy density 
of CRs at the source will be much larger than that needed of the 
excess electrons due to the fact that the diffusion time of PeV CRs 
is shorter than the energy loss time of TeV electrons in propagation.}. 
It can be seen that the interaction between CR nuclei and background 
photons naturally bridges the knee of CR spectra and the excesses of
electrons/positrons.

It is generally believed that the Galactic CRs are accelerated by
the shock waves in supernova remnants (SNRs). We will focus on the
acceleration of CRs by the neutron star inside the shell of young 
SNRs for our purpose, because that is where abundant high energy CRs 
(projectile) and background photons (target) can be found. As described 
in \cite{1987Natur.329..314G,1987Natur.329..807B,1989ApJ...345..423G},
the CRs up to ultra high energies (e.g., $10^{18}$ eV) can be accelerated 
in a very short period ($\sim$yr or shorter) after the explosion of the
supernova. The radiation field has an initial temperature of about
several thousand K, which is equivalent to a photon energy of
$\sim$eV. Before the radiation field decays and the temperature
cools down in hundreds of days, the e$^+$e$^-$ pair production
process can occur many times for one CR particle. As both the
magnetic field and radiation field are very strong in the case of young
SNRs, the electron energy can not exceed a few TeV because of the
synchrotron depletion and the inverse Compton scattering. It is
worth noting that there should also be energy gain for electrons
through acceleration. As shown in \cite{2009A&A...497...17V} the
electron acceleration in radiation dominated environments should
exhibit a pile up spectrum around the cutoff energy, which should
be a few TeV in the case of young SNRs \citep{1990cup..book.....G}.
The properties of young SNRs seem to fit well with what is required 
to understand the knee of CR spectra and the abnormal spectra of
electrons/positrons, but a quantitative study is necessary.

For simplicity we decouple the acceleration and interaction processes
in the calculation, i.e., the nuclei are first accelerated to very
high energies and then they interact in the photon field. Finally,
we inject the nuclei and electrons/positrons in the Galaxy to calculate
the propagation effect and derive the observational spectra on the
Earth. Monte-Carlo (MC) method is used to calculate the energy losses of
the nuclei and the energies of the generated e$^+$e$^-$ pairs.

There are three processes of interactions between CRs and photons,
i.e., pair production, photodisintegration and photo-pion
production. The cross sections for pair production and
photodisintegration are given in \cite{1970PhRvD...1.1596B}
and \cite{1976ApJ...205..638P} respectively. The pion production cross
section for proton is adopted from \cite{2008PhLB..667....1P}, and
we employ an $A^{0.91}$ dependence for other nuclei with atomic number
$A$\citep{1985PhRvD..32.1244S}.

We choose a power-law spectrum with index about $2$ for primary CRs,
according to the Fermi acceleration mechanism \citep{1990cup..book.....G}.
The actual spectral index of the three main components, proton, Helium and
Iron, are slightly tuned in the calculation to match the data. For other
nuclei which do not play a significant role we adopt the parameterization
$\gamma_Z=a+bZ-0.6$ with the fitting parameters $a=2.69$ and $b=-2.07\times
10^{-3}$ \citep{2003APh....19..193H}. Here the $0.6$ subtraction from
parameter $a$ is done to explicitly take into account the propagation
correction from the observations to the source spectra (see below).
The relative abundances of various kinds of nuclei after
propagation are normalized to the fluxes at $1$ TeV given in
\cite{2003APh....19..193H}. The propagation effects of nuclei are
approached using a leaky-box model. The escape time for CR nuclei is
taken as $\tau_{\rm esc}(R)\approx 2\times 10^8\left(\frac{R}
{1\,{\rm GV}}\right)^{-0.6} {\rm yr}$, where $R=p/Ze$ is the rigidity
of nuclei. This relation is fitted from the low energy B/C data
\citep{2008arXiv0808.2437P} and we extrapolate it directly to the PeV
energy range.

\begin{figure*}[!htb]
\centering
\includegraphics[width=0.45\columnwidth]{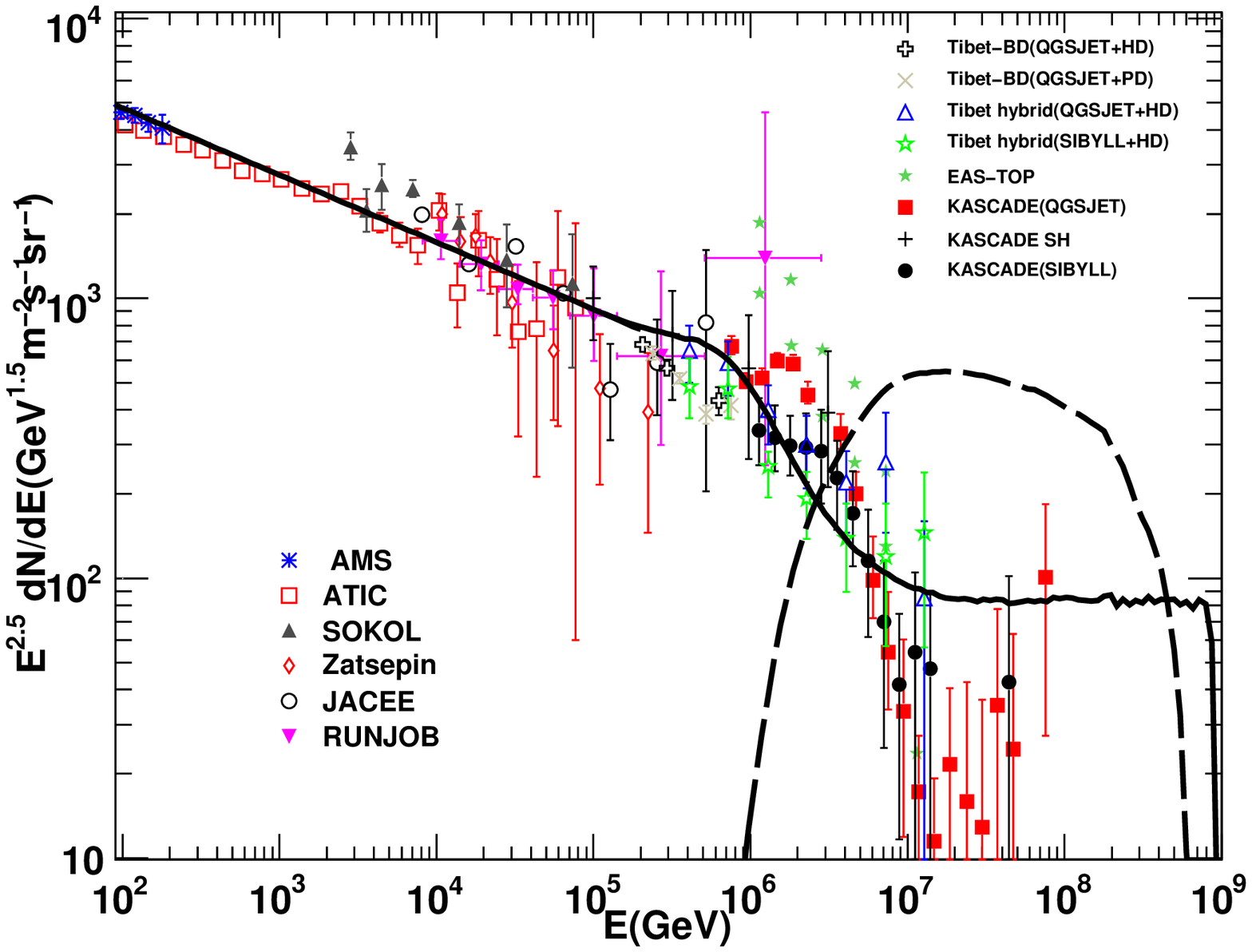}
\includegraphics[width=0.45\columnwidth]{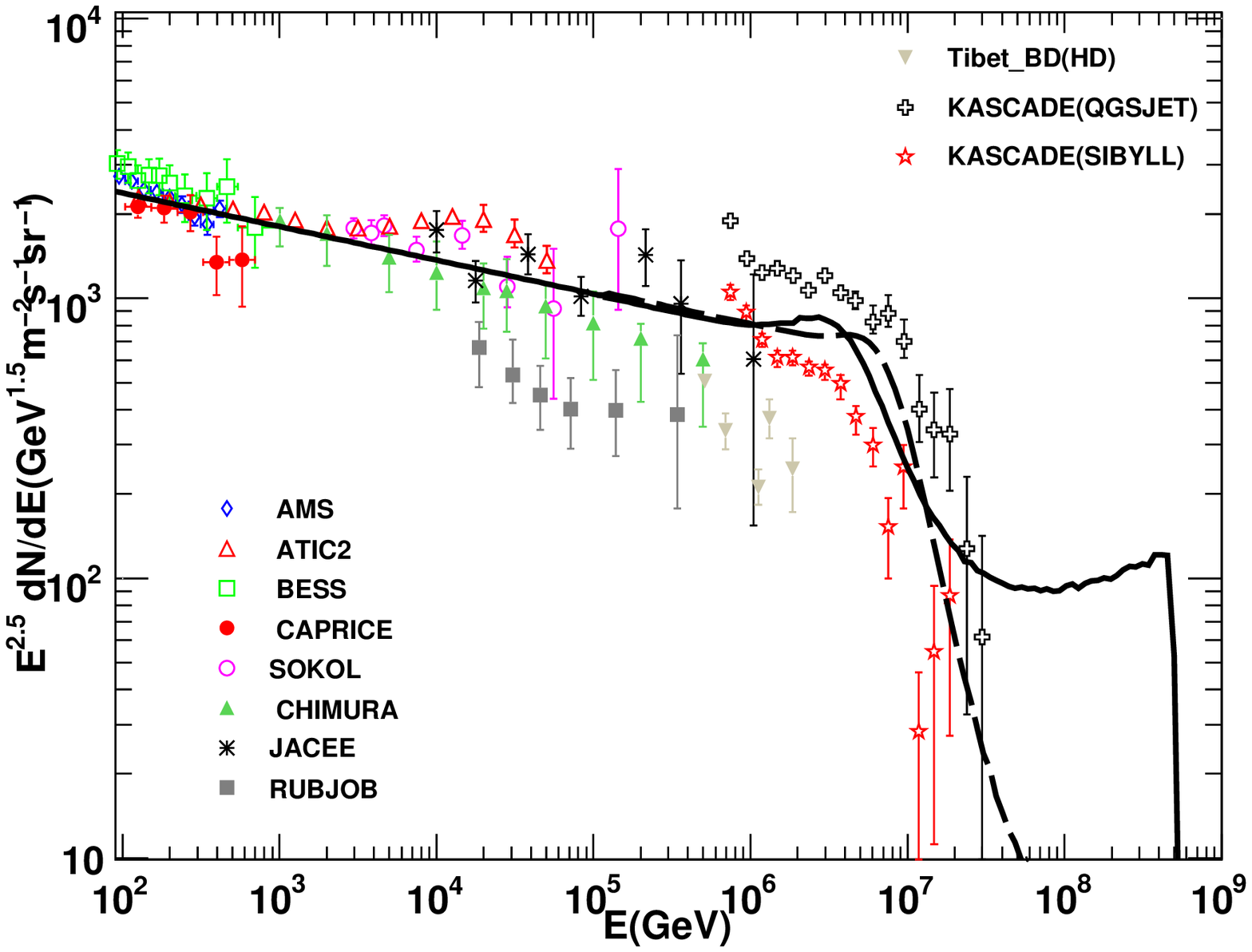}
\includegraphics[width=0.45\columnwidth]{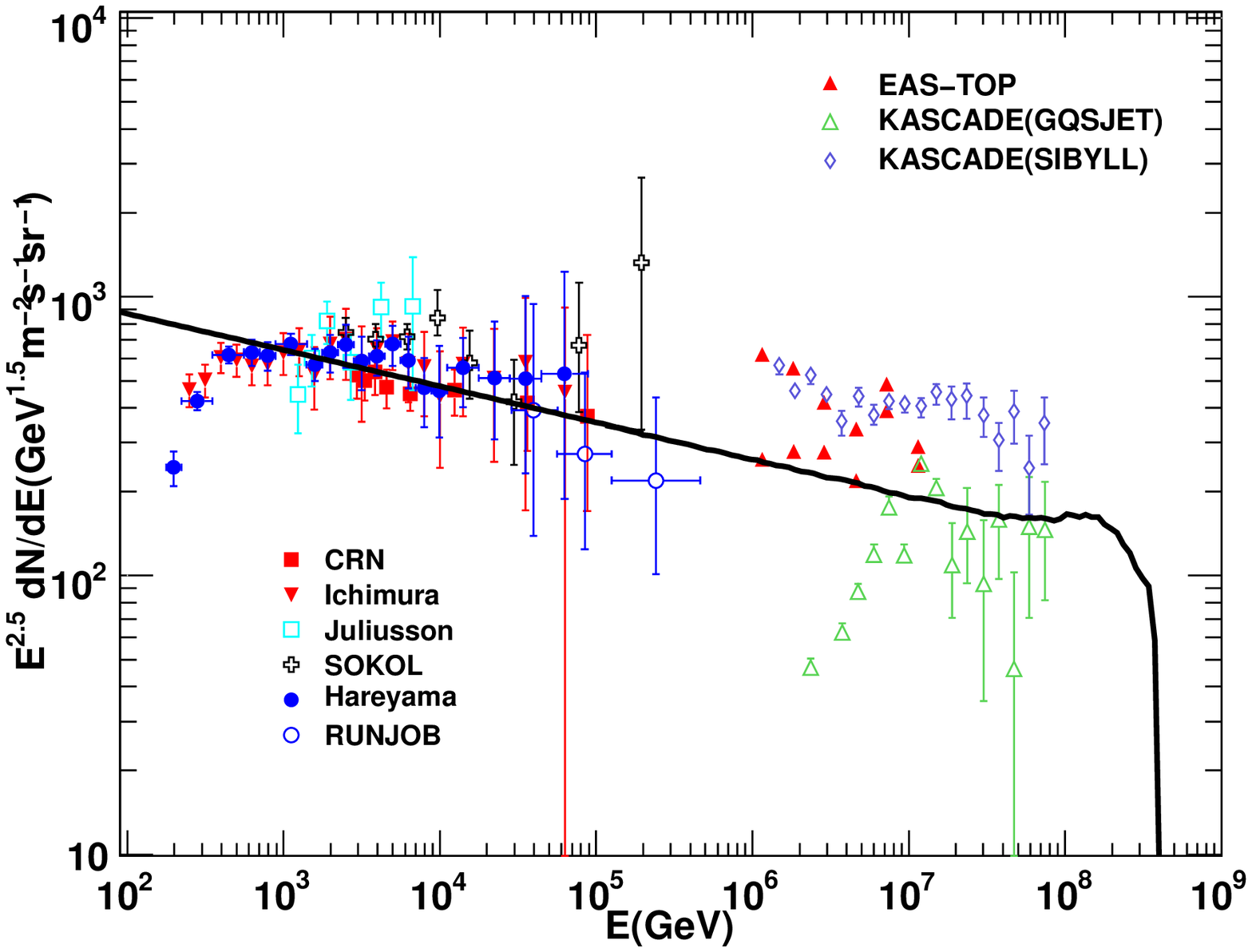}
\includegraphics[width=0.45\columnwidth]{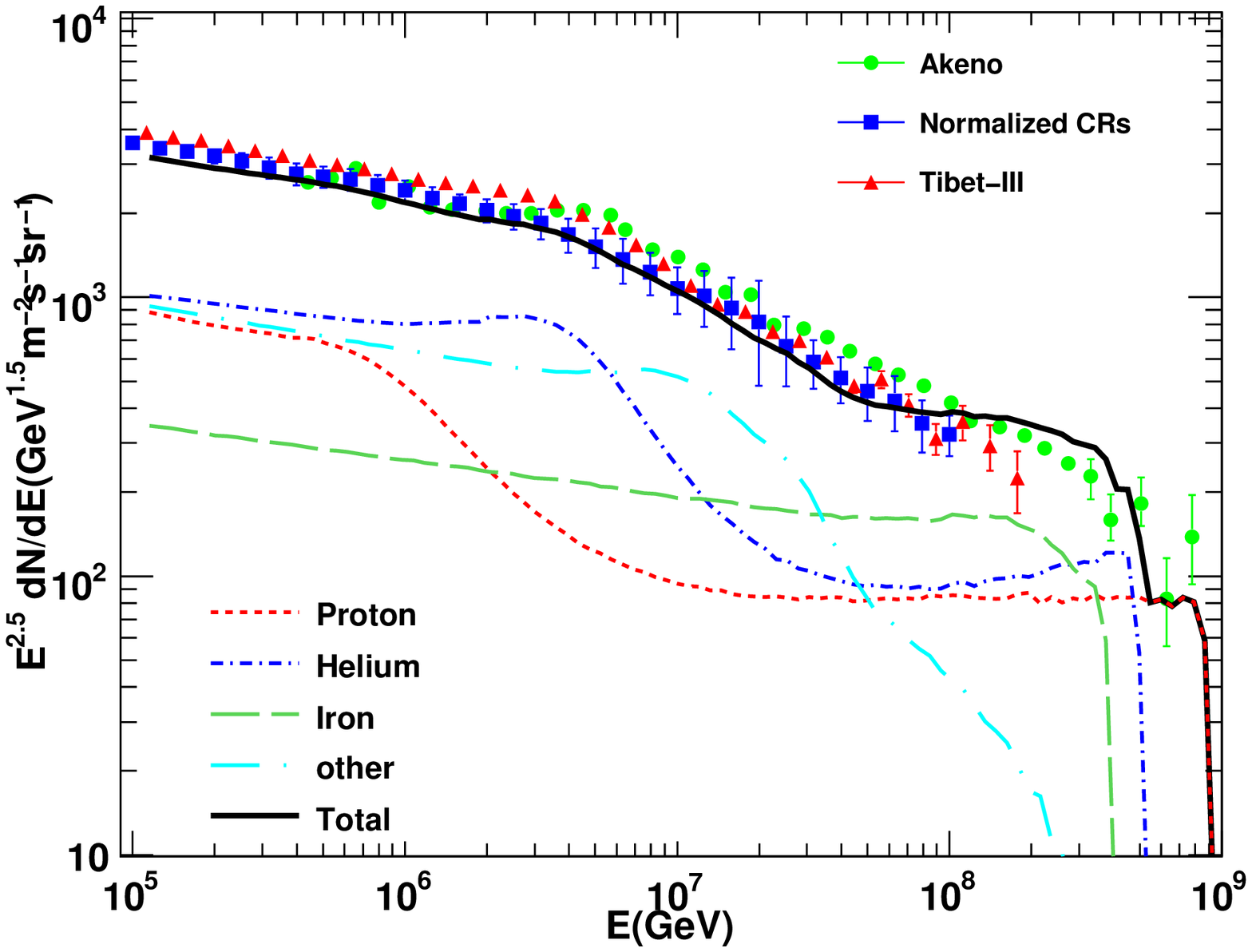}
\caption{\small
Solid lines in each figure are the calculated energy spectra
for protons, Helium, Iron and total CRs respectively, after including
the pair production interactions with photons. The dashed lines
in the upper two panels are Helium spectrum with pair production,
photodisintegration and pion production, and the secondary proton
spectrum due to photodisintegration of Helium.
Observational data are, proton: Tibet hybrid \citep{2006JPhCS..47...51A},
Tibet-BD \citep{2000PhRvD..62k2002A}, JACEE \citep{1998ApJ...502..278A},
RUNJOB \citep{2001APh....16...13R}, KASCADE \citep{2005APh....24....1A,
2004ApJ...612..914A}, EAS-TOP \citep{2003ICRC....1..147N},
ATIC \citep{2003ICRC....4.1833A}, Zatsepin \citep{1993ICRC....2...13Z},
SOKOL \citep{1993ICRC....2...17I}, AMS \citep{2000PhLB..490...27A};
Helium: JACEE \citep{1998ApJ...502..278A}, KASCADE \citep{2005APh....24....1A},
Tibet-BD \citep{2000PhRvD..62k2002A}, ATIC2 \citep{2006astro.ph.12377P},
RUNJOB \citep{2001APh....16...13R}, SOKOL \citep{1993ICRC....2...17I},
Ichimura \citep{1993PhRvD..48.1949I}, BESS \citep{2004PhLB..594...35H},
AMS \citep{2002PhR...366..331A}, CAPRICE \citep{2003APh....19..583B};
Iron: EAS-TOP \citep{2003ICRC....1..147N}, KASCADE \citep{2005APh....24....1A},
Juliusson \citep{1974ApJ...191..331J}, Ichimura \citep{1993PhRvD..48.1949I},
CRN \citep{1991ApJ...374..356M}, Hareyama \citep{1999ICRC....3..105H},
RUNJOB \citep{2001APh....16...13R}, SOKOL \citep{1993ICRC....2...17I};
total: Tibet-III \citep{2008ApJ...678.1165A}, Akeno \citep{1984JPhG...10.1295N}.
The normalized data are derived by combining all data with a rescale based
on the extrapolation of the direct measurements \citep{2003APh....19..193H}.
}
\label{picture2}
\end{figure*}

As for the radiation field around the source, the temperature evolution 
with time is a crucial factor to define the dominant type of interactions 
between CRs and background photons. In a simple case, the temperature
remains unchanged when CRs are being accelerated to ultra high energies. 
The three types of interactions may then happen simultaneously. However 
this may not be true providing that the
threshold energies of photodisintegration and pion production are
much higher than that of pair production. The pair production
should start earlier and slows down the acceleration of CRs,
which makes the other two processes with higher thresholds more
difficult to occur. The observational data actually support this
argument. As shown by the dashed lines in the upper two panels of
Fig.\ref{picture2}, if we allow the photodisintegration to be
active when fitting the Helium spectrum, we find that protons are
over produced and the calculated proton spectrum does not agree
with the data. The suppression of photodisintegration can be understood 
by considering that the radiation field decays and/or cools down rapidly
\citep{1987Natur.329..314G,
1987Natur.329..807B,1989ApJ...345..423G} compared to the time scale
of the CR acceleration. This leaves only the pair production
as the main interaction between CRs and radiation photons.
Therefore in the following calculation we forbid the
photodisintegration and pion production and only consider pair
production.

Based on the previous considerations, the temperature and effective photon
density are fitted to the observed spectra of individual elements. As shown by
the solid lines in Fig.\ref{picture2}, the calculated spectra of proton,
Helium, Iron and the all particle CRs agree well with the measurements.
In Table \ref{table1}, the parameters used in the simulation are summarized: 
the temperature, the effective column density of photon and interaction
time assumed in a blackbody radiation field, and the initial power law index
of each element. In this work we assume the second knee ($\sim 300$PeV)
of CR spectra is due to the interaction between Iron and background photons
but this assumption is not a mandatory one (see below the discussion on
systematic check). The temperatures and photon densities of other nuclei
are derived using a linear interpolation with respect to the atomic
number $A$ between Helium and Iron. The heavy nuclei (low charge over
mass ratio) correspond to low temperatures because they need longer time
to be accelerated to reach the pair production threshold when the 
temperature of the source has decreased. It can also be seen from Table 
\ref{table1} that if the photon field is as intense as a blackbody 
radiation, an interaction time from a few weeks to a few months is enough 
to produce the observed spectrum distortion.

\begin{figure}[!htb]
\centering
\includegraphics[width=\columnwidth]{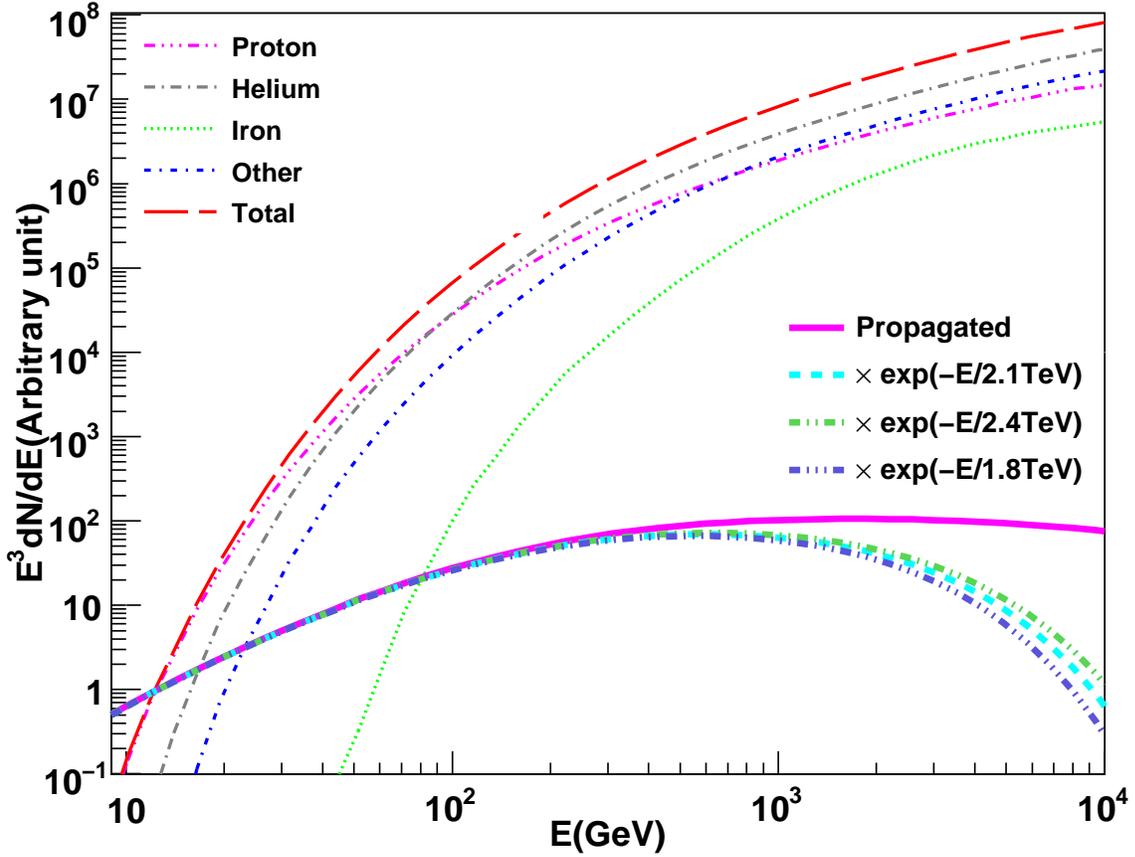}
\caption{{\it Upper-thin lines:} Electron/positron spectra generated
through photon-nuclei interactions for various CR components.
{\it Lower-thick lines:} Propagated electron/positron spectra with
arbitrary normalization. Also shown are the results with an exponential
cutoff of the injection spectra with $E_c=1.8,\,2.1,\,2.4$ TeV
respectively \citep{2008PhRvL.101z1104A}.}
\label{elec_spec}
\end{figure}

The generated spectra of electrons/positrons at the source location for
individual nuclei and all CRs are shown in Fig.\ref{elec_spec}. 
The propagation effect of electrons/positrons in the Galaxy has to be 
considered to determine the spectrum on Earth. Unlike CRs,
the dominant effect for electron propagation is the energy losses due
to inverse Compton scattering and synchrotron radiation at energies 
higher than $\sim 10$GeV. In this work the GALPROP code
\citep{1998ApJ...509..212S} is used to calculate the propagation of
electrons and positrons.

\begin{figure*}[!htb]
\centering
\includegraphics[width=0.45\columnwidth]{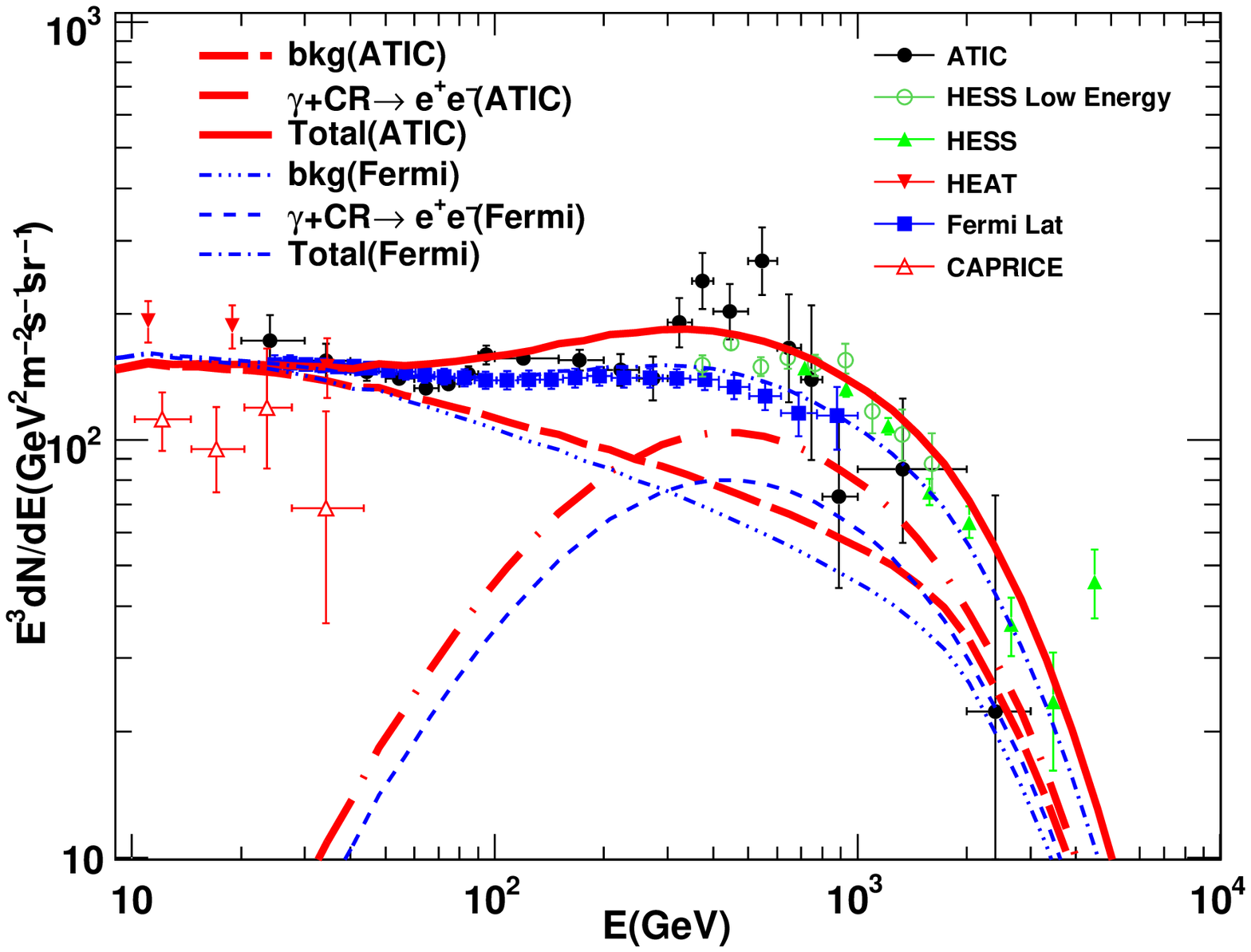}
\includegraphics[width=0.45\columnwidth]{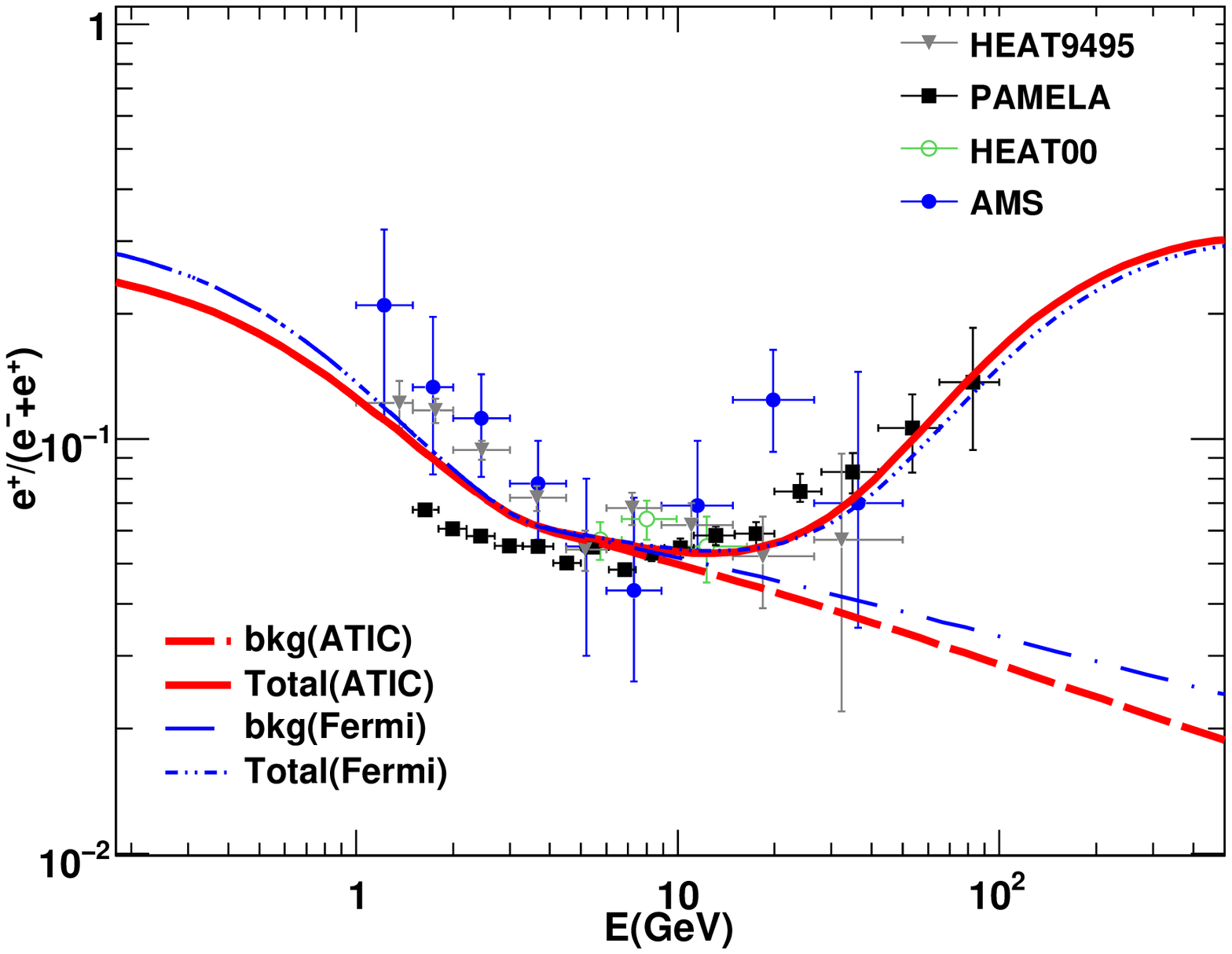}
\caption{{\it Left:} Total spectra for e$^++$e$^-$. The labels ``ATIC''
and ``Fermi'' in the brackets mean the results to fit ATIC or Fermi data 
respectively. Observational data are: CAPRICE \citep{2000ApJ...532..653B},
HEAT \citep{1998ApJ...498..779B}, ATIC \citep{2008Natur.456..362C},
H.E.S.S. \citep{2008PhRvL.101z1104A}, H.E.S.S. low energy
\citep{Aharonian:2009ah}, Fermi-LAT \citep{Fermi:2009zk}.
{\it Right:} Positron fraction. Observational data are:
AMS \citep{2007PhLB..646..145A}, HEAT94+95 \citep{1997ApJ...482L.191B},
HEAT00 \citep{2001ICRC....5.1687C}, PAMELA \citep{2009Natur.458..607A}.
}
\label{atic_pamela}
\end{figure*}

\begin{table}
\centering
\caption{Parameter settings in the Monte-Carlo calculation.}
\begin{tabular}{ccccc}
\hline \hline   & $T_{\rm ph}$ & $\langle nc\tau\rangle$ & $\hat{\tau}$ & $\gamma_Z$ \vspace{-0mm} \\
  & (K) & ($10^{29}$cm$^{-2}$) & (yr) &  \\
\hline
  Proton & $1\times 10^4$ & $8.1$ & $0.04$ & $2.14$ \\
  Helium & $7\times 10^3$ & $12.9$ & $0.19$ & $2.02$ \\
  Iron   & $2\times 10^3$ & $0.9$ & $0.58$ & $2.03$ \\
  \hline
  \hline
\end{tabular}
\label{table1}
\end{table}

As shown by the lower thick lines in Fig.\ref{elec_spec}, the propagated
electron spectrum becomes softer due to the cooling effects from energy
losses. It is however still harder than the experimental data. But as
pointed out in \cite{1989ApJ...345..423G}, the high energy tail 
($\gtrsim$TeV) of electrons could be depleted in the environment of young 
SNR by the strong magnetic and radiation fields. An accurate calculation
needs to involve a detailed modeling of radiation, interaction and 
acceleration. As this work focuses on the interaction rather than the
acceleration and radiation of electrons, we will simply adopt an exponential
cutoff on the injection spectra of electrons/positrons with a value
experimentally suggested by H.E.S.S. measurements
\citep{2008PhRvL.101z1104A}. The low curves in Fig.\ref{elec_spec} show
the effects of different values of cutoff energy on the electron spectrum.

The calculated (e$^++$e$^-$) spectrum and the positron fraction for a 
cutoff energy $E_c\approx 2$ TeV are shown by the thick lines in 
Fig.\ref{atic_pamela} together with the observational data. The 
calculated flux of (e$^++$e$^-$) is scaled to match the data by a 
free normalization factor (quantitatively, it is $\sim 3\%$ for this
model configuration).  This parameter represents the dissipation and 
escape probability of electrons/positrons in the source region. In
Fig.\ref{atic_pamela}, the term ``bkg'' represents the standard
background contribution of primary electrons and secondary
electrons/positrons from CR interactions with the interstellar
medium. For the background calculation, we use the diffusion +
convection (DC) model \citep{2009PhRvD..79b3512Y} with parameters
tuned to be consistent with the low energy CR data. For background
electrons at the high energy end ($\sim$TeV) we implement the same
exponential cutoff as described above. The results show good
agreement with the observational data by e.g., ATIC and PAMELA.
Furthermore, by slightly adjusting the background and a $20\%$ 
decrease of the generated $e^++e^-$ flux, the Fermi, H.E.S.S. and 
PAMELA data can also be well fitted simultaneously, as shown by the 
thin lines in Fig.\ref{atic_pamela}.

As a systematic check, while still assuming that pair production is 
the dominant interaction between CRs and photons, a constant source 
temperature is tried to reproduce the observed spectra. We find that 
the calculated spectra are also in agreement with the observational 
data for temperatures from $5000$ K to $7000$ K.

In summary we propose to explain the features of the CR spectra by the
e$^+$e$^-$ pair production at the acceleration sources in an environment
similar to the young SNRs. We show that the spectra of CRs and 
electrons/positrons agree well with the observations. This in turn provides 
a strong support that at least a portion of CRs might be accelerated at 
young SNRs. Our work provides a new interpretation to understand the 
abnormal positron fraction observed by PAMELA and the electrons excesses 
by ATIC/Fermi in the standard astrophysical framework. In addition, we 
can note in our model that only one simple set of the physical parameters 
of the CR sources seem to be sufficient to give good descriptions to the 
major properties of the data. It may indicate: 1) all the sources are 
``standard'' and have similar parameters, 2) the average effect is 
equivalent to using single set of parameters, 3) one single nearby source 
dominates the observed fluxes of CRs, and so on. If the CRs indeed
come from a single nearby source\footnote{Note, in such a case the 
background calculation of electrons/positrons using GALPROP might not
be proper any more. However, phenomenologically we can adjust the source 
parameters to get similar spectra of electrons/positrons for a single 
source with the continous source distribution, although there might 
be some constraints from, e.g., the diffuse gamma-rays. To build a
fully self-consistent model of a single source scenario is beyond the 
scope of the present study.}, one would expect a sharp knee of the 
CR spectra \citep{1997JPhG...23..979E}. It is interesting to note that
the Tibet AS$\gamma$ experiment indeed observed a sharp knee structure
\citep{2008ApJ...678.1165A}. Furthermore, the young SNR acceleration 
model predicts a sharp cutoff of electron spectra beyond several TeV 
which can be tested by future high quality experiments, such as the 
upgraded Tibet AS$\gamma$ experiment with underground MUON detectors 
\citep{2008JPhCS.120f2024A}, 
GRAPES\footnote{http://alpha.sci.osaka-cu.ac.jp/grapes3/index.html} and
KASCADE\footnote{http://www-ik.fzk.de/KASCADE\_home.html}. Finally, 
we note that the current measurements of CR spectra around the knee 
energy are of great uncertainties for individual CR nuclei. If precise 
measurements become available in the future, it will be very helpful 
in understanding the model and in determining the model parameters.

\acknowledgments

We thank Li Tipei, Ding Linkai, L\"u Caidian, Wang Jianxiong, Li Haibo,
Liu Siming, Deng Jinsong, Lu Fangjun, Zhu Shouhua, Lou Yuqing, Olivier
Martineau-Huynh, Artin Teymourian and Wang Hanguo for helpful discussions. 
Hu Hong-Bo thanks Wang Wei for 
the long term help of the research work. This work is supported by the 
Minister of Science and Technology of China, Natural Sciences Foundation 
of China (Nos. 10725524 and 10773011), and by the Chinese Academy of
Sciences (Nos. KJCX2-YW-N13 and KJCX3-SYW-N2).

\end{document}